\documentclass[12pt]{article}
\usepackage[
backend=biber,
style=ieee,
minbibnames=6,
maxbibnames=6,
doi=false,
isbn=false,
url=false,
eprint=false,
sorting=none,
]{biblatex}
\usepackage[colorlinks=true,linkcolor=blue]{hyperref}
\usepackage{geometry}
\usepackage{times}

\usepackage[utf8]{inputenc}
\usepackage{amssymb}
\usepackage{float}

\usepackage{amsmath}
\usepackage{graphicx}
\usepackage{multirow}
\usepackage{threeparttable}

\usepackage{comment}
\usepackage{array}
\usepackage{colortbl}
\usepackage{xcolor}
\usepackage{arydshln}
\usepackage{makecell}
\usepackage{afterpage}
\newcolumntype{L}[1]{>{\raggedright\let\newline\\\arraybackslash\hspace{0pt}}m{#1}}
\newcolumntype{C}[1]{>{\centering\let\newline\\\arraybackslash\hspace{0pt}}m{#1}}
\newcolumntype{R}[1]{>{\raggedleft\let\newline\\\arraybackslash\hspace{0pt}}m{#1}}
\usepackage{tabularx}

\title{Music Similarity Representation Learning Focusing on Individual Instruments with Source Separation and Human Preference}

\author{Takehiro Imamura, Yuka Hashizume, Wen-Chin Huang, Tomoki Toda\\Nagoya University, Aichi, Japan}

\begin{document}
\maketitle

\begin{abstract}
This paper proposes music similarity representation learning (MSRL) based on individual instrument sounds (InMSRL) utilizing music source separation (MSS) and human preference without requiring clean instrument sounds during inference.
We propose three methods that effectively improve performance.
First, we introduce end-to-end fine-tuning (E2E-FT) for the \textit{Cascade} approach that sequentially performs MSS and music similarity feature extraction.
E2E-FT allows the model to minimize the adverse effects of a separation error on the feature extraction.
Second, we propose multi-task learning for the \textit{Direct} approach that directly extracts disentangled music similarity features using a single music similarity feature extractor.
Multi-task learning, which is based on the disentangled music similarity feature extraction and MSS based on reconstruction with disentangled music similarity features, further enhances instrument feature disentanglement.
Third, we employ perception-aware fine-tuning (PAFT).
PAFT utilizes human preference, allowing the model to perform InMSRL aligned with human perceptual similarity.
We conduct experimental evaluations and demonstrate that 1) E2E-FT for \textit{Cascade} significantly improves InMSRL performance, 2) the multi-task learning for \textit{Direct} is also helpful to improve disentanglement performance in the feature extraction, 3) PAFT significantly enhances the perceptual InMSRL performance, and 4) \textit{Cascade} with E2E-FT and PAFT outperforms \textit{Direct} with the multi-task learning and PAFT.
\end{abstract}

\section{Introduction}\label{sec:introduction}
Recently, the number of musical pieces available online has already exceeded 1 billion\footnote{\url{https://go.pardot.com/l/52662/2023-10-23/ljk7xt/52662/169805013966KGzgtB/Spotify_2023_Culture_Next_Report_JP_v3.pdf}} and futher market expantion is expected.\footnote{\url{https://www.ifpi.org/wp-content/uploads/2020/03/Global_Music_Report_2023_State_of_the_Industry.pdf}}
Therefore, the demand for music recommendation and retrieval systems has been increasing.
Methods utilizing listening histories of users~\cite{collaborate1, collaborate2} have been widely used in these systems although these methods cause several limitations, for example, it is hard to handle musical pieces with less listening records.
One approach to avoid this problem is to extract content features from a musical piece and utilize them for music recommendation and retrieval.
Music feature extraction models using classical methods \cite{classical1, classical2, classical3}, such as Mel-Frequency Cepstrum Coefficients \cite{classical_mfcc}, K-Means \cite{classical_kmeans}, Gaussian Mixture Modeling \cite{classical_mfccgmm1, classical_mfccgmm2}, fluctuation patterns \cite{pam1, fp1}, etc, have been investigated.
Recently, methods based on deep learning have attracted attention due to their high precision of music feature extraction.
Many studies utilize a Convolutional Neural Network (CNN) \cite{cnn1, cnn2, cnn3, cnn4, cnn5, musicnn} and have demonstrated their high performance on their tasks.

In paticular, music similarity representation learning~(MSRL) with unsupervised, self-supervised or semi-supervised learning methods have gained popularity since it can handle previously unseen data and can be applied to various downstream tasks. 
Aritists labels \cite{artist_label} or music tags \cite{words} have been used for training with triplet loss \cite{triplet_loss}, and their positive impact on a downstream music classification task or zero-shot performance has been demonstrated.
Futhermore, contrastive learning approaches that assume ``segments within the same song are similar to each other'' \cite{s4, mule}, which is called S4 in this paper, or utilize data augmentation \cite{clmr, mule_da} have demonstrated the strong effectiveness on several tasks \cite{mabble}, despite not requiring any labels.
Moreover, HuBERT-style \cite{hubert} masked language modeling \cite{mlm}, which estimates the tokens corresponding to masked parts of the input, has also achieved outstanding effectiveness on many downstream tasks \cite{mabble} by utilizing teacher labels obtained from K-Means clurstering \cite{mert}, the EnCodec model \cite{encodec} or the data2vec-styled \cite{data2vec} approach \cite{music2vec}.
Additionally, approaches utilizing the intermediate layer outputs of music generation models \cite{jukebox, jukebox_tl} and CLIP-style \cite{clip} text-music contrastive learning methods \cite{mulan} have been established as techniques for acquiring music representations.

Although MSRL extracts features that can be applied to various downstream tasks, it performs only a single feature extraction per musical piece. 
For recommendation and retrieval applications, which need to account for individual users' preference differences, it is desirable to obtain multiple features from diverse perspectives.
MSRL based on individual instrument sounds~(InMSRL)~\cite{hashizume_inst, hashizume_mix} is a potential technique to develop a new function allowing users to focus on multiple partial elements of musical pieces, e.g., searching for musical pieces with similar drum sounds.
Conventional InMSRL models have been trained with the similarity assumption S4, which has also been demonstrated to be effective in MSRL \cite{s4, mule}.
This approach is advantageous in that it does not require any labels for training, especially since such labels for individual instrument sounds are rarely available.
To further obtain the music similarity representation for each instrument from the musical pieces that multiple instrument sounds are mixed, we have proposed three main approaches: \textit{Clean} \cite{hashizume_inst}, \textit{Cascade} \cite{hashizume_inst}, and \textit{Direct} \cite{hashizume_mix}.
\textit{Clean}, which inputs clean individual instrument sounds into the corresponding music similarity feature extractors, have been demonstrated its high performance in the music similarity feature extraction per each instrument sounds.
While \textit{Clean} requires clean individual instrument sounds as a searching query during inference, such sounds are generally not publicly available, making it practically impossible to utilize it in general-purpose music recommendation and retrieval systems.
Therefore, research on the InMSRL model that can input musical pieces themselves during inference and accurately obtain the music similarity representation per individual instrument sounds has progressed, leading to the proposal of \textit{Cascade} and \textit{Direct}.
\textit{Cascade} sequentially performs music source separation~(MSS) \cite{spleeter} and music similarity feature extractions.
With this feature extraction strategy, the model is expected to clearly disentangle the individual instrument features from musical pieces.
On the other hand, \textit{Direct} extracts disentangled music similarity features using a single music similarity feature extractor, with the goal of learning a disentangled feature space consisting of different subspaces for the individual instrument sounds, similar to \cite{veit2017conditional, lee2020disentangled}.
With this architecture, the model can reduce the computational costs for feature extraction compared to \textit{Cascade} and avoid artifacts from the preceding model (e.g., MSS models in \textit{Cascade}) in feature extraction.

However, in \textit{Cascade}, since the MSS model and music similarity feature extractors are independently trained, separation errors are likely to cause adverse effects on the music similarity feature extraction.
Furthermore, in \textit{Direct}, learning disentangled feature representation is not straightforward, and InMSRL performance tends to degrade for certain instruments.
Additionally, although \textit{Clean}, \textit{Cascade}, and \textit{Direct} employs the S4-based training and successfully earn the music similarity representation between the same musical pieces, there is no guarantee that this training approach captures similarity between different musical pieces—or corresponds to human perceptual similarity.
Indeed, the previous study \cite{abx} have shown that while the music similarity representation from S4-based models between the same musical pieces has exhibited a strong correlation with human perception, the music similarity representation from models between different musical pieces has shown an insufficient correlation with human perception.

In this paper, we propose InMSRL methods utilizing music source separation and human preference aiming to construct a universally applicable InMSRL model and acquire a music similarity representation reflecting human perceptual similarity.
For \textit{Cascade}, we propose \textit{Cascade-FT} that performs end-to-end fine-tuning~(E2E-FT) of the pre-trained MSS model and the music similarity feature extractors using an auxiliary separation loss.
For \textit{Direct}, we propose \textit{Direct-Reconst} that uses multi-task learning based on the disentangled music similarity feature extraction and MSS based on reconstruction~(Reconst) with the disentangled music similarity features.
Furthermore, to allow the model to perform InMSRL aligned with human perceptual similarity, we introduce perception-aware fine-tuning (PAFT) utilizing a small amount of human preference labels.
We conduct experimental evaluations and demonstrate that 1) the E2E-FT for \textit{Cascade} significantly improves InMSRL performance, 2) the multi-task learning for \textit{Direct} is also helpful to improve disentanglement performance in the feature extraction, 3) PAFT significantly enhances the perceptual InMSRL performance, and 4) \textit{Cascade} with the E2E-FT and PAFT outperforms \textit{Direct} with the multi-task learning and PAFT.

The rest of this paper is organized as follows: 
In Section \ref{sec: conv_inmsrl}, we describe the previously proposed InMSRL methods. 
In Section \ref{sec: proposed inmsrl}, we provide the details of our proposed method.
In Section \ref{sec: ex_eval}, we evaluate the proposed method through experimental evaluation.
Finally, in Section \ref{sec: conclusion}, we present our conclusion.

\section{Conventional InMSRL Methods}\label{sec: conv_inmsrl}
\subsection{\textit{Clean} and \textit{Cascade}}\label{sec: related_clean_cascade}
Hashizume et al.~\cite{hashizume_inst} proposed two InMSRL methods: one inputting clean individual instrument sounds into the corresponding music similarity feature extractors (\textit{Clean}) and the other inputting individual instrument sounds separated by the pre-trained MSS model of Spleeter~\cite{spleeter} into those extractors (\textit{Cascade}).

The music similarity feature extractors of \textit{Clean} and \textit{Cascade} are trained using a triplet loss. 
In the $i$-th triplet, three types of sample segments, anchor~$\mathbf{x}_i^{(a)}$ that serves as the basis, positive~$\mathbf{x}_i^{(p)}$ defined as similar to the anchor and negative~$\mathbf{x}_i^{(n)}$ defined as dissimilar to the anchor, are used.
By denoting a distance function as $d(\cdot)$, a loss function can be formulated as follows:
\begin{equation}
    \label{eq:related_triplet}
    \displaystyle
    \begin{split}
    \mathcal{L}_\mathrm{triplet}
    = \mathrm{max}\{0, d(\mathbf{x}_i^{(a)}, \mathbf{x}_i^{(p)}) - d(\mathbf{x}_i^{(a)}, \mathbf{x}_i^{(n)}) + \delta\}\\            
    \end{split}
\end{equation}
where $\delta$ is a margin that defines the minimum distance between the anchor-positive and anchor-negative pairs.
To perform label-free learning, assuming S4 condition, a triplet is constructed as follows:
\begin{itemize}
    \item Anchor: Extracted from a randomly selected musical piece
    \item Positive: Extracted from the same musical piece as that of the anchor 
    \item Negative: Extracted from a different musical piece from that of the anchor.
\end{itemize}

In \textit{Cascade}, it is inevitable to cause separation errors in MSS.
The previous studies~\cite{hashizume_inst} have confirmed that the performance of \textit{Cascade} significantly degrades compared with \textit{Clean}.
Therefore, it is crucial to optimize the MSS model for the instrument-dependent music similarity feature extractors.

\subsection{\textit{Direct}}\label{sec: related direct}
Hashizume et al. \cite{hashizume_mix} also proposed the other InMSRL method to extract a disentangled music similarity feature with a single feature extractor, where the disentangled music similarity feature consists of subspaces for individual instrument-dependent music similarity features, e.g., the first to 128-th dimensional components of the 640-dimensional feature vector are used to represent the music similarity focusing on drums. 

The training process first involves pre-training. 
In this training, the single disentangled music similarity feature extractor is trained using a target disentangled feature formed by concatenating the instrument-dependent music similarity features extracted by \textit{Clean}.

Next, similar to \textit{Cascade}, the disentangled music similarity feature extractor is further updated by using the triplet loss as shown in Eq. \ref{eq:related_triplet}.
However, unlike \textit{Cascade}, it is not straightforward to train such a feature extractor. To develop the disentangled music similarity feature extractor working reasonably, the following two approaches are used.
\begin{itemize}
    \item Conditioning the output of the disentangled music similarity feature extractor
    \item Using pseudo-musical-pieces as inputs.
\end{itemize}

Conditioning process conduct a masking operation inspired by other disentangled representation learning methods \cite{veit2017conditional, lee2020disentangled}.
For example, when focusing on the bass feature, we leave only the dimensional components corresponding to a subspace for the bass feature and masks the other dimensional components to 0.
By partially masking the feature vector, each subspace can model the music similarity feature depending on a specific instrument sound.

\begin{figure}[t]
    \centerline{
        \includegraphics[clip,  width=12cm]
        {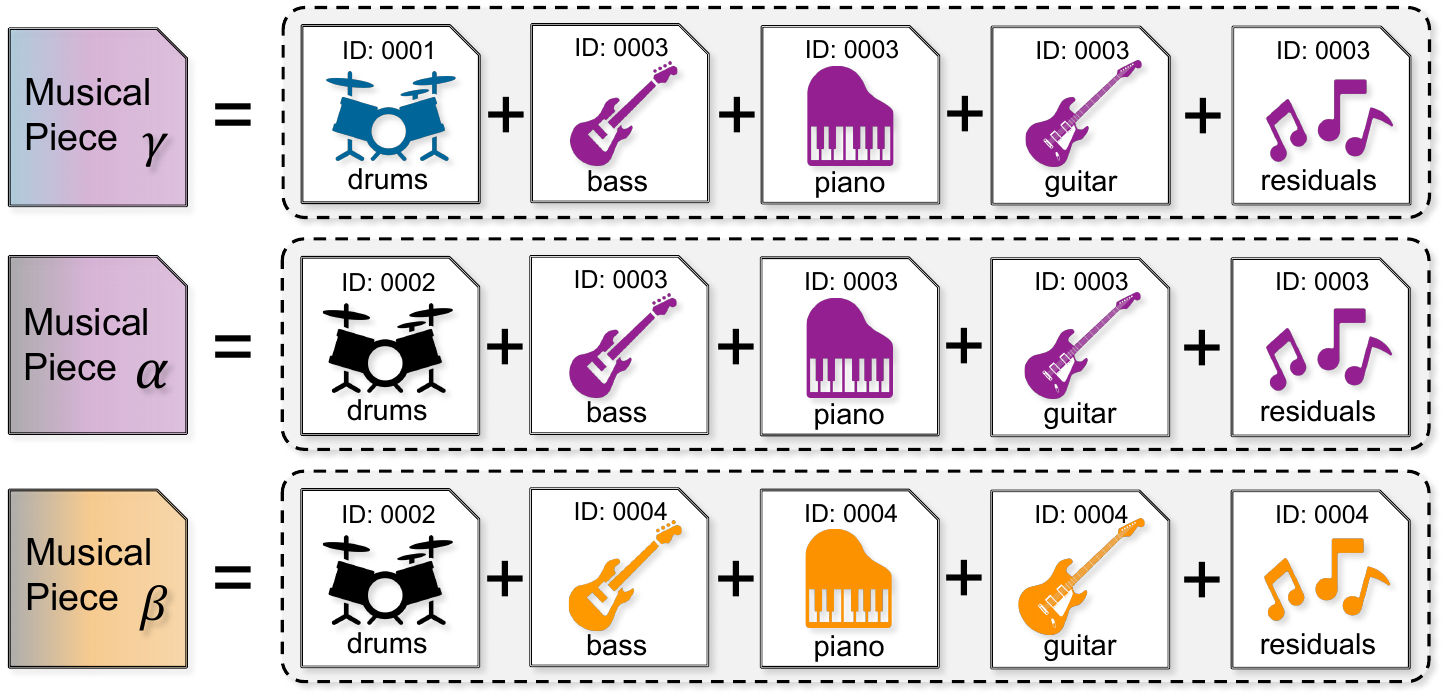}
        }
    \caption{Overview of pseudo-musical-pieces. Instruments of the same color and the same ID indicate sample segments extracted from the same musical piece. This figure illustrates an example of the pseudo-musical-pieces created for learning focusing on drums.}
    \label{fig:pseudo_song}
\end{figure}

The use of pseudo-musical-pieces aims to improve the disentangled performance of \textit{Direct}.
Fig.~\ref{fig:pseudo_song} shows an overview of the pseudo-musical-pieces.
In Fig.~\ref{fig:pseudo_song}, a musical piece~$\alpha$ and a musical piece~$\beta$ are similar to each other in drums but dissimilar in the other instruments sounds. In contrast, the musical piece~$\alpha$ and a musical piece~$\gamma$ are dissimilar in drums but similar in the other instruments.
In the triplet loss-based learning, by using the musical piece~$\alpha$ as the anchor, the musical piece~$\beta$ as the positive, and the musical piece~$\gamma$ as the negative, the model can focus only on the drum features.
By treating this triplet setting as the basic triplet data, a previous study \cite{hashizume_mix} further introduced additional triplet data. 
Specifically, the additional triplet data are constructed by swapping the positive and negative samples in the basic triplet data, and, during training with the additional triplet data, a different instrument from that in the basic triplet data is targeted.

However, it is still challenging to accurately disentangle a musical piece into the instrument-dependent subspace features. Consequently, the performance of InMSRL based on \textit{Direct} tends to be insufficient.

\subsection{Perceptual Music Similarity Representation Performance of an Conventional Method}\label{sec: related abx}
Hashizume et al. have collected a large-scale dataset of human preference labels and have analyzed the human perception of similarity between individual instrumental sounds within musical pieces \cite{abx}. 
Specifically, they have conducted an ABX test with 586 participants, where each participant was asked to answer the question: ``Which of A or B is more similar to X?'', given three segments of musical pieces (X, A, and B).
In constructing the ABX dataset, they defined the following two conditions for comparison:
\begin{itemize}
    \item All-Diff: X, A, and B are extracted from entirely different songs
    \item One-Shared: Either A or B is extracted from the same song as X
\end{itemize}
The ABX dataset included drums, bass, piano, guitar, residuals, and mix tracks, with 240 pairs for each condition, totaling 480 pairs of ABX data. 
Here, residuals refer to all sounds in a musical piece except for drums, bass, piano, and guitar, while mix represents the full audio mix of the musical piece. 
Each pair of ABX data was evaluated by at least three participants, resulting in a total of 26,898 valid responses.

Furthremore, an experimental evaluation about the perceptual InMSRL performace of the conventional InMSRL model trained in S4 condition \cite{hashizume_mix} have been conducted in \cite{abx}.
The evaluation results showed that for the One-Shared condition, there was a strong correlation between human perceptual music similarity and music similarity by the S4-based model. 
Since, in the One-Shared condition, either A or B can satisfy S4 assumption, it is inferred that the S4 criterion aligns to some extent with human perceptual music similarity, and the conventional InMSRL model can represent the similarity between the same musical piece.
However, in the All-Diff condtion, only a weak correlation have been observed between the human perceptual music similarity and the music similarity by the S4-based model.
This indicates that the conventional InMSRL model is inadequate to represent the music similarity between segments from different musical piece.
Given that S4-based training successfully captures similarity aligned with S4, it is possible that this approach is not effective in learning similarity between different musical pieces that do not conform to the S4 criterion.

\section{Proposed InMSRL Methods Leveraging Multi-Task Learning and Human Preference}\label{sec: proposed inmsrl}
\subsection{\textit{Cascade-FT}}
To address the issue of \textit{Cascade}, we propose \textit{Cascade-FT} to optimize the MSS model by performing end-to-end fine-tuning (E2E-FT).

\subsubsection{Network Architecture}
\begin{figure}[t]
    \centering
        \includegraphics[width=1.0\columnwidth]{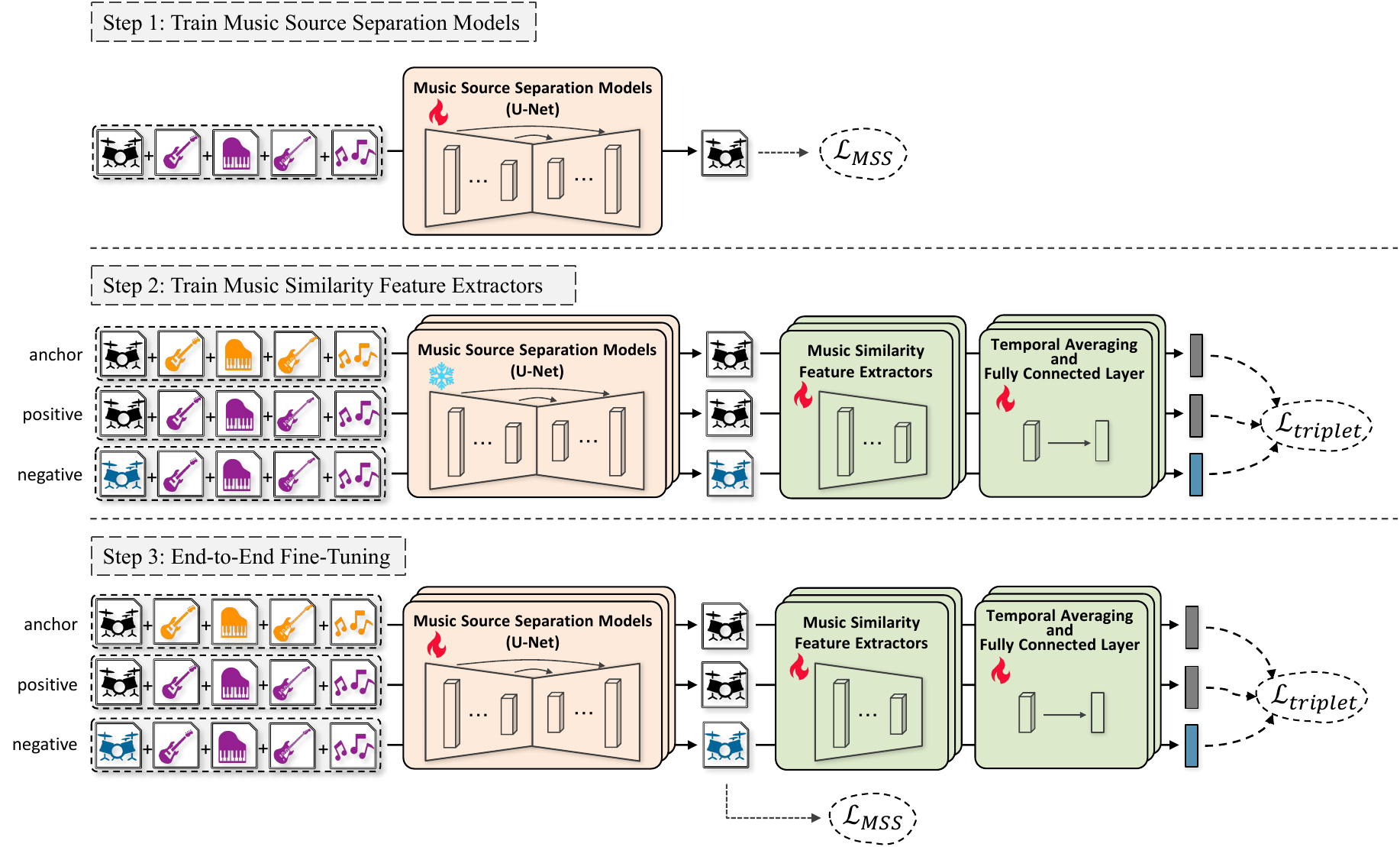}
    \caption{Overview of \textit{Cascade-FT} model.}
    \label{fig:proposed_inst}
\end{figure}

The network architecture of \textit{Cascade-FT} consists of the MSS model and the instrument-dependent music similarity feature extractor connected in series as shown in Fig.~\ref{fig:proposed_inst}.
The MSS model is based on the U-Net~\cite{2015unet, singing_voice_separation} structure, similar to the Spleeter~\cite{spleeter} used in \textit{Cascade}.
The network outputs a separation mask and the separated instrument sound is generated by Hadamard product of the input music spectrogram and the separation mask.
In this paper, we develop the instrument-dependent MSS models to separately estimate the separation masks for individual instrument sounds.
The instrument-dependent music similarity feature extractor is based on the U-Net encoder structure additionally using time-averaging and flattening operations and a fully-connected layer to output a 128-dimensional feature vector for each instrument sound.

\subsubsection{Training}\label{sec: cascade-ft_training}
The training procedure consists of three stages: training of the MSS models, training of the instrument-dependent music similarity feature extractors and E2E-FT.
First, the MSS models are trained in the same manner as proposed by Jansson et al.~\cite{singing_voice_separation}. 
The separation loss for each instrument sound (denoted as $\mathcal{L}_{MSS}$ in Fig.~\ref{fig:proposed_inst}) is calculated as the L1 loss between the output separated instrument sound and a clean target instrument sound.
Next, the music similarity feature extractors are trained using the triplet loss given by Eq. \ref{eq:related_triplet} (denoted as $\mathcal{L}_{triplet}$ in Fig.~\ref{fig:proposed_inst}) in the same manner as in \textit{Cascade}. 
During the training, the MSS models are frozen and their parameters are not updated. 
The L2 norm is employed as the distance function $d(\cdot)$ in the triplet loss.
Finally, in the E2E-FT stage, all parameters of the cascaded network consisting of the MSS models and the instrument-dependent music similarity feature extractors are updated by using a combined loss function given by the triplet loss for the instrument-dependent music similarity feature extractors and the separation loss for the MSS models as an auxiliary loss.
Note that three inputs (anchor, positive, and negative) are required to compute the triplet loss, the auxiliary separation loss for the MSS models during fine-tuning is calculated for all three inputs.
In the training, the pseudo-musical-pieces segments (shown in Fig.~\ref{fig:pseudo_song}) are also used as in \textit{Direct}. Besides, we implement the data augmentation as described in Section \ref{sec:direct-reconst_dt}.


\subsection{\textit{Direct-Reconst}}
To address the issue of \textit{Direct}, we propose \textit{Direct-Reconst} incorporating MSS based on the reconstruction~(Reconst) with the disentangled music similarity features for the training of the disentangled feature extractors.

\subsubsection{Network Architecture}
\begin{figure}[t]
    \centering
        \includegraphics[width=1.0\columnwidth]{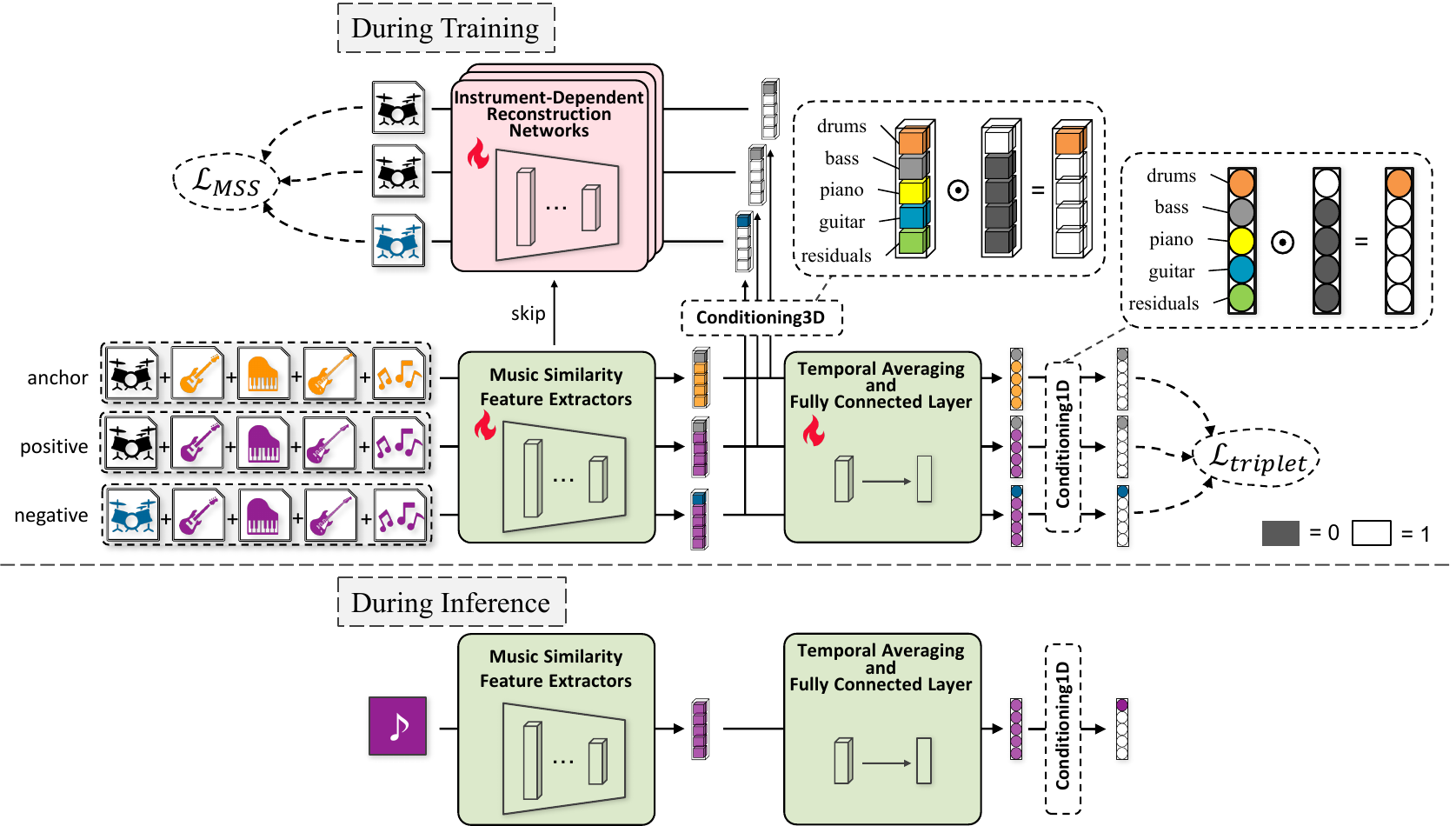}
    \caption{Overview of \textit{Direct-Reconst} model. The same color of inputs and outputs of the networks indicate the segments extracted from the same musical pieces.}
    \label{fig:proposed_mix}
\end{figure}

Fig.~\ref{fig:proposed_mix} shows the network architecture of \textit{Direct-Reconst}.
The \textit{Direct-Reconst} network consists of three parts: the disentangled music similarity feature extractor, a reconstruction network to reconstruct each instrument sound from output sequences of the disentangled music similarity feature extractor, and a time-averaging and flattening operations and fully-connected layer to generate the disentangled music similarity feature vector from the output sequences.
The disentangled music similarity feature extractor has a similar structure to the encoder of U-Net~\cite{singing_voice_separation}, and the reconstruction network has a similar structure to the decoder of U-Net~\cite{singing_voice_separation}. 
The each layer of the disentangled music similarity feature extractor and those of the reconstruction network are connected by skip connections.
The instrument-dependent reconstruction networks are developed for individual instrument sounds.
As in the MSS models, the reconstructed instrument sound is generated by Hadamard product of the input music source spectrogram and the output separation mask.

\subsubsection{Training}\label{sec: direct-reconst_training}
The training procedure consists of two stages: pre-training of the music similarity feature extractor and multi-task learning of the music similarity feature extractor and the instrument-dependent reconstruction network.
In the pre-training of the music similarity feature extractor, we follow the same training procedure as in \textit{Direct}~\cite{hashizume_mix}. 
We use 31 out of the $2^5$ possible combinations of 5 musical instrument sources (drums, bass, piano, guitar, and residuals) as input, excluding the silent pattern.
The training loss for the multi-task learning is a combination of the triplet loss given by Eq. \ref{eq:related_triplet} (denoted as $\mathcal{L}_{triplet}$ in Fig.~\ref{fig:proposed_mix}) for the disentangled music similarity features and the reconstruction loss (denoted as $\mathcal{L}_{MSS}$ in Fig.~\ref{fig:proposed_mix}) for the output reconstructed instrument sounds.
The distance function $d(\cdot)$ in the triplet loss for the disentangled music similarity features is defined as the L2 norm.
The reconstruction loss is defined as the L1 loss between the output instrument sound from the reconstruction network and the clean instrument sound in the same manner proposed by Jansson et al.~\cite{singing_voice_separation}.
As in \textit{Direct}, we use the conditioning operation and the pseudo-musical-pieces.

\subsubsection{Disentanglement Enhancement}\label{sec:direct-reconst_dt}
To enhance the disentangled music similarity feature extractor, we modify the conditioning process and utilize pseudo-musical-pieces.
The modified conditioning process applies the masking operation to not only the output of the time-averaging and flattening operations and fully-connected layer~(Conditioning1D in Fig.~\ref{fig:proposed_mix}) but also the input of the reconstruction network~(Conditioning3D in Fig.~\ref{fig:proposed_mix}).
Conditioning1D is the same as the conditioning process used in \textit{Direct}.
Conditioning3D is its extension to apply the masking operation to a feature sequence.
By Conditioning3D, the reconstruction network can focus only on the features corresponding to each target instrument sound.
For the pseudo-musical-pieces, we further introduce data augmentation (DA).
While \textit{Direct} generates a fixed set of triplet data of the pseudo-musical-pieces beforehand and use it in the training, \textit{Direct-Reconst} introduces a process of randomly generating triplet data of the pseudo-musical-pieces each time to construct a mini-batch during training.

\subsection{Fine-Tuning Utilizing Human Preference}
For improving the perceptual InMSRL performance, we propose a \\Perception-Aware Fine-Tuning (PAFT) which utilizes few human preference labels obtained from ABX test \cite{abx} as described in Section \ref{sec: related abx}.
The training with PAFT follows a two-step process.
First, the training with triplet loss of each InMSRL model (described in Section \ref{sec: related_clean_cascade}, \ref{sec: related direct}, \ref{sec: cascade-ft_training}, and \ref{sec: direct-reconst_training}) is conducted as pre-training.
Then, PAFT fine-tunes these models using the human preference data obtained from ABX test.
For the loss function for PAFT, the triplet loss given by Eq. \ref{eq:related_triplet} is used.
The data setting for the triplet loss is defined as follows:
\begin{itemize}
    \item Anchor: The reference data in ABX test (denoted as X in Section \ref{sec: related abx})
    \item Positive: The segment (A or B) that is determined to be more similar to X in the ABX data
    \item Negative: The segment (A or B) that is determined to be less similar to X in ABX data
\end{itemize}
For \textit{Clean}, the clean individual instrument sounds are used for the inputs of models during PAFT.
For \textit{Cascade} and \textit{Direct} approach, the pseudo-musical-pieces are used during PAFT.
Here, the pseudo-musical-pieces during PAFT are defined using only the ABX data for the target instrument, while the non-target instrument sounds are defined in the same manner as described in Section \ref{sec: related direct}.
During PAFT of \textit{Cascade}, \textit{Cascade-FT}, \textit{Direct}, and \textit{Direct-Reconst}, only the feature extractors are trained and the other parts of the model are frozened.
With PAFT, the model can aquire the music similarity representation that aligns with human perceptual music similarity.

\section{Experimental Evaluations}\label{sec: ex_eval}
\subsection{Dataset}
The dataset used for evaluation was the Slakh~\cite{slakh}, which was also used in the previous study~\cite{hashizume_inst, hashizume_mix}. The dataset consisted of MIDI-generated musical pieces and MIDI instrument tracks in the musical pieces.
Each musical piece contained drums, bass, piano, and guitar sounds. Following previous studies~\cite{hashizume_inst, hashizume_mix}, the other sounds in the musical pieces were treated as residuals.

The dataset consisted of 2100 musical pieces containing multiple groups of musical pieces generated from the same MIDI file. In this experiment, we excluded musical pieces generated from the same MIDI file, resulting in 1200 musical pieces used for training, 270 musical pieces used for validation, and 136 musical pieces used for evaluation.

\subsection{Evaluation Metrics}
As an evaluation metric, we used music ID estimation accuracy and perceptual similarity agreement as used in the previous studies \cite{hashizume_inst, hashizume_mix, abx}. 
Here, for music ID estimation accuracy, we used the following two metrics, a music ID estimation score on normal-test-musical-pieces (MES-Normal) and a music ID estimation score on pseudo-test-musical-pieces (MES-Pseudo).

\subsubsection{Music Estimation Score on Normal-Test-Musical-Pieces (MES-Normal)}
To evaluate the performance of the feature representation, we used the accuracy of the music ID estimation with a simple method using the feature representation.
Specifically, assuming that all test segments were embedded into the learned feature space beforehand and the music IDs of all segments were known except for a test segment to be estimated,
we used the 5-nearest neighbors (5NN) method to estimate the music ID of the test segment. 
In the evaluation for each instrument sound, 
we only used feature dimensions corresponding to the target instrument in 5NN distance calculation while masking the other feature dimensions.
The entire test dataset~(136 musical pieces) was used to calculate the music ID estimation accuracy.


\subsubsection{Music Estimation Score on Pseudo-Test-Musical-Pieces (MES-Pseudo)}
\begin{figure}[t]
    \centering
        \includegraphics[clip,  width=14cm]
        {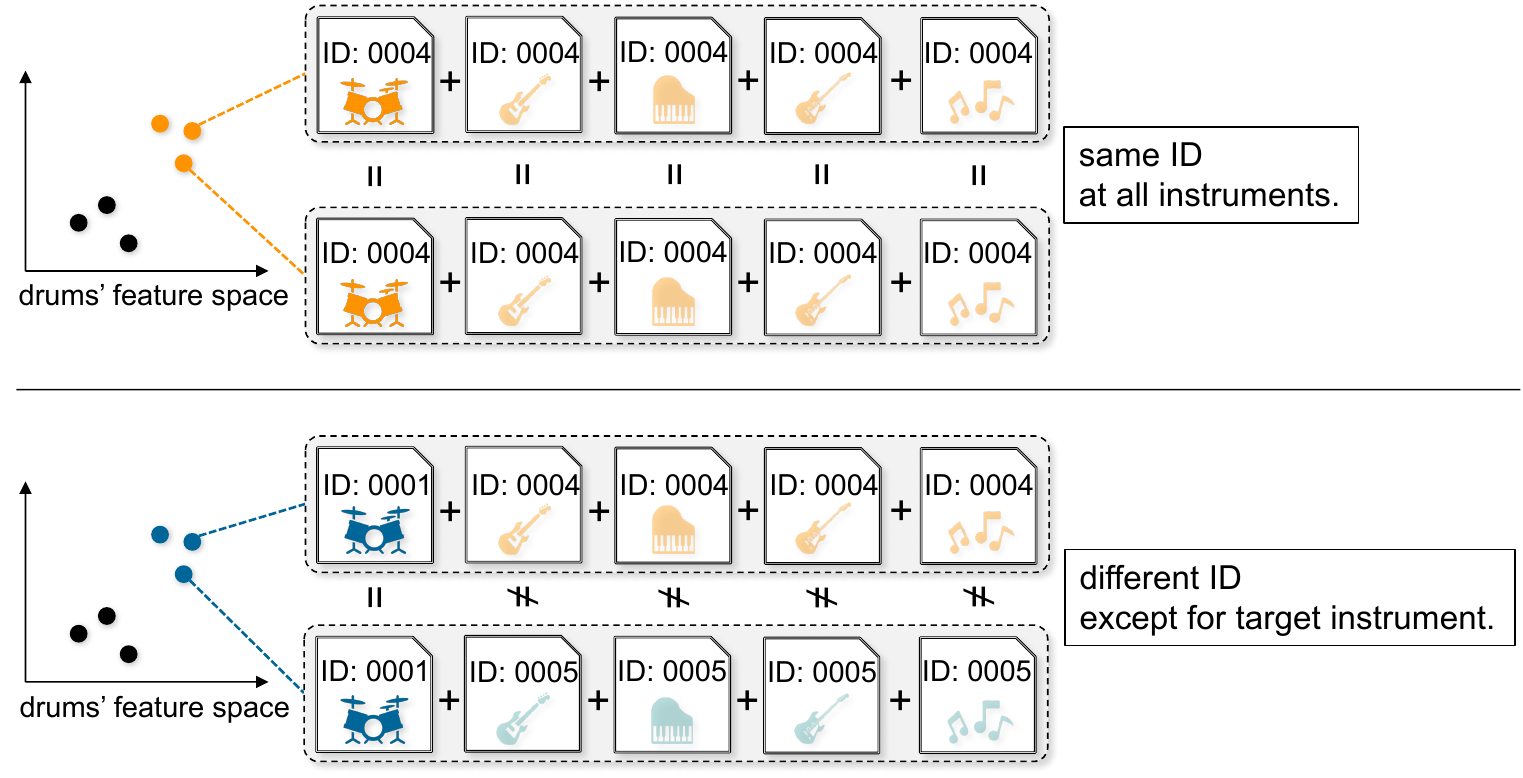}
    \caption{Difference between MES-Normal and MES-Pseudo. The top part of the figure shows MES-Normal, and the bottom part shows MES-Pseudo. This is the example of evaluation for the drums. Instruments of the same color and the same ID indicate segments extracted from the same musical piece.}
    \label{fig:normal_pseudo}
\end{figure}

The proposed method aimed to learn the music similarity feature representations focusing on individual instrument sounds.
However, in MES-Normal, the ground truth label for the 5NN method was the same over all instrument sounds as shown in the top part of Fig.~\ref{fig:normal_pseudo}.
Therefore, it was essentially hard to evaluate the disentanglement performance of the learned representations by MES-Normal.
To investigate the disentanglement performance, we used pseudo-test-musical-pieces in MES-Pseudo. 
In MES-Pseudo, the ground truth label was different between the target instrument and the others; e.g., the label of the target instrument sound (i.e., drums label) was different from the others as shown in the bottom part of Fig.~\ref{fig:normal_pseudo}.
The pseudo-musical-pieces used for the test were generated as follows: 
1) 10 musical pieces were selected from the dataset to be used for the target instrument sounds, 
2) those 10 musical pieces were removed from the dataset, 
3) for each of those 10 target musical pieces, 3 musical pieces were further selected from the remaining dataset, and they were used for the non-target instrument sounds, 
and 4) each of the 10 target musical pieces and the corresponding 3 non-target musical pieces were mixed to generate 30 pseudo-musical-pieces in total. 
The entire test dataset was constructed by using those 30 pseudo-musical pieces as well as the 10 normal musical pieces used for the target musical pieces, consisting of 40 musical pieces in total. 
In the test, we excluded all segments extracted from the same musical-piece as that of each test segment to prevent the music ID estimation focusing on the nontarget instruments; i.e., when using a segment within one of the 30 pseudo-musical pieces as a test segment, only segments within 2 pseudo-musical pieces and 1 normal musical piece had correct music ID labels; on the other hand, when using a segment within one of the 10 normal musical pieces as a test segment, only segments within 3 pseudo-musical pieces had correct musical ID labels.

\subsubsection{Visualization of Music Similarity Feature Vectors}
To gain a deeper evaluation of the performance of each InMSRL method, we introduced the visualization of music similarity features.
During visualization, pseudo-musical-pieces were used as inputs of models.
The used pseudo-musical-pieces were constructed as follows. 
First, 10 musical pieces were selected. 
They were used for both the target instrument and non-target instruments to determine the pseudo musical pieces. 
Consequently, 100 pairs of the target musical piece and the non-target musical pieces were used in total to generate the pseudo musical pieces. 
Next, 10 segments were retrieved from each musical piece. 
Finally, 10 pseudo-segments were constructed for each of those 100 pseudo-musical-pieces by randomly selecting target and non-target segments from those retrieved 10 segments of the corresponding musical pieces and mixing them. 
Note that although 10 out of those 100 pseudo musical pieces were equivalent to the normal musical pieces, the constructed pseudo-segments were usually different from normal ones because of these mixing process using the randomly selected segments. 
In total, 1,000 pseudo-segments were used to visually investigate which instrument sounds the model focused on.

\subsubsection{Perceptual Similarity Agreement}
To evaluate the perceptual InMSRL performance of models, we used perceptual similarity agreement utilizing ABX data obtained from previous research \cite{abx}.
Specifically, the accuracy was calculated by comparing human responses and model predictions for the question, ``Which of A or B is more similar to X?'', given three segments of musical pieces (X, A, and B).
The details of ABX data are described in Section \ref{sec: related abx}.
The ABX data used for evaluation included only the data where more than 75\% of participants consistently selected one of the two options, either A or B, as the segment most similar to the reference segment X.
The number of ABX data samples and subject responses are shown in Table \ref{tbl: abx_data}.

\begin{table}[tb]
    \centering
    \caption{The number of ABX data and subject responses.}
    \label{tbl: abx_data}
    \scalebox{0.88}[0.88]{
        \begin{tabular}{l|cccc}
            \hline
            Method & 
            drums & 
            bass & 
            piano & 
            guitar
            \\
            
            \hline

            \rowcolor{lightgray!50} The number of ABX data &&&&\\

            
            all data (All-Diff) & 
            240 & 
            240 & 
            240 &
            240
            \\

            all data (One-Shared) & 
            240 & 
            240 & 
            240 &
            240
            \\

            above 75\% (All-Diff) & 
            115 & 
            112 & 
            106 &
            105
            \\

            above 75\% (One-Shared) & 
            214 & 
            194 & 
            215 &
            213
            \\

            \hline

            \rowcolor{lightgray!50} The number of subject responses &&&&\\


            all (All-Diff) & 
            2421 & 
            2429 & 
            2424 &
            2425
            \\

            all (One-Shared) & 
            2408 & 
            2422 & 
            2425 &
            2412
            \\

            above 75\% (All-Diff) & 
            1163 & 
            1128 & 
            1083 &
            1085
            \\

            above 75\% (One-Shared) & 
            2163 & 
            2959 & 
            2159 &
            2134
            \\
            
            \hline
        \end{tabular}
    }
\end{table}

\subsection{Experimental conditions}
Music segments used in the experiments were cut into 3-second segments for training based on S4 similarity, 5-second for PAFT and perceptual similarity agreement, and 10-second segments for validating and MES-Normal and MES-Pseudo.
The training data and test data for PAFT were split from the ABX data obtained in ABX test \cite{abx} at a $7:3$ ratio.
The music segments where the target instrument was silent were excluded.
The sampling rate was set to 44,100 Hz, and a window size of 2,048 and a frame shift of 512 were used for the short-time Fourier transform~(STFT).
The number of mel frequencies for the log Mel-spectrogram used as input to the \textit{Cascade-FT} music similarity feature extractors was set to 259.
The learning rate in \textit{Cascade-FT} was set to $5 \times 10^{-5}$ for training based on S4, $1 \times 10^{-5}$ for fine-tuning, and $5 \times 10^{-5}$ for PAFT.
The learning rate for the pre-training of \textit{Direct-Reconst} and the multi-task training of the disentangled music similarity feature extractor and the reconstruction network was set to $1 \times 10^{-4}$.
Adam \cite{kingma2017adam} was used to train both models. 
The maximum number of epochs was set to 400 except for PAFT, and training was terminated if the minimum value of the loss function on the validation data was not updated over 100 epochs.
In PAFT, the number of epochs was set to 100.

\subsection{Experimental Results}
\begin{table}[tb]
    \centering
    \caption{Evaluation results of MES-Normal~(\%). The evaluation scores of \textit{Clean}~\cite{hashizume_inst}, \textit{Cascade} w/ Spleeter~\cite{hashizume_inst} and \textit{Direct}~\cite{hashizume_mix} are respectively quoted from~\cite{hashizume_inst} and~\cite{hashizume_mix}. In w/o pseudo-musical-pieces, the music similarity feature extractors are simply trained with normal musical pieces. Excluding ablation, the best results are highlighted in bold. ``\textcolor{gray}{Gray}'' text indicates the scores used in the ablation study.}
    \label{table:ex_result_normal}
    \scalebox{0.8}[0.8]{
        \begin{tabular}{l||ccccc}
            \hline
            Method 
            & drums 
            & bass 
            & piano 
            & guitar 
            & residuals
            \\
            
            \hline
            
            \textit{Clean}~\cite{hashizume_inst} 
            & 98.04     
            & 94.60     
            & 98.14     
            & 96.35 
            & -
            \\
            
            \hdashline
            
            \textit{Cascade} w/ Spleeter~\cite{hashizume_inst} 
            & 88.91     
            & 63.87 
            & 50.34     
            & - 
            & -
            \\
            
            \textit{Cascade} w/ from-scratch 
            & 90.98     
            & 73.39     
            & 80.77     
            & 79.53 
            & -
            \\

            ~~~~~~\textcolor{gray}{w/o pseudo-musical-pieces}
            & \textcolor{gray}{92.71} 
            & \textcolor{gray}{90.20} 
            & \textcolor{gray}{93.62} 
            & \textcolor{gray}{90.90} 
            & \textcolor{gray}{-}
            \\
            
            ~~~w/ E2E-FT~(\textit{Cascade-FT}) 
            & \bf 93.03 
            & \bf 74.96     
            & \bf 81.96 
            & \bf 82.78 
            & -
            \\
            
            ~~~~~~\textcolor{gray}{w/o pseudo-musical-pieces} 
            & \textcolor{gray}{94.89} 
            & \textcolor{gray}{95.63} 
            & \textcolor{gray}{96.21} 
            & \textcolor{gray}{94.40} 
            & \textcolor{gray}{-}
            \\
            
            \hdashline

            \textit{Direct}~\cite{hashizume_mix} 
            & 89.69     
            & \bf 84.45 
            & \bf 85.70 
            & \bf 86.27 
            & 84.86
            \\

            ~~~w/ DA
            & 89.33     
            & 71.09     
            & 79.74     
            & 81.75     
            & 85.67
            \\
            
            ~~~w/ DA, Reconst~(\textit{Direct-Reconst}) 
            & \bf 91.14 
            & 81.30     
            & 84.76     
            & 85.17 
            & \bf 88.84
            \\
            
            \hline
        \end{tabular}
        }
\end{table}
\begin{table}[tb]
    \centering
    \caption{Evaluation results of MES-Pseudo~(\%). The evaluation scores of \textit{Direct}~\cite{hashizume_mix} are quoted from the previous study \cite{hashizume_inst}. In w/o pseudo-musical-pieces, the music similarity feature extractors are simply trained with normal musical pieces. Excluding ablation, the best results are highlighted in bold.``\textcolor{gray}{Gray}'' text indicates the scores used in the ablation study.}
    \label{table:ex_result_pseudo}
    \scalebox{0.8}[0.8]{
        \begin{tabular}{l||ccccc}
            \hline
            Method 
            & drums 
            & bass 
            & piano 
            & guitar 
            & residuals
            \\ 
            
            \hline
            
            \textit{Cascade} w/ from-scratch 
            & 98.68     
            & 93.02     
            & 91.73     
            & 92.19 
            & -
            \\
            
            ~~~~~~\textcolor{gray}{w/o pseudo-musical-pieces} 
            & \textcolor{gray}{95.09} 
            & \textcolor{gray}{77.30} 
            & \textcolor{gray}{81.02} 
            & \textcolor{gray}{77.60} 
            & \textcolor{gray}{-}
            \\
            
           ~~~w/ E2E-FT~(\textit{Cascade-FT}) 
            & \bf 98.91 
            & \bf 94.80 
            & \bf 93.55 
            & \bf 93.89 
            & -
            \\
            
            ~~~~~~\textcolor{gray}{w/o pseudo-musical-pieces} 
            & \textcolor{gray}{95.96} 
            & \textcolor{gray}{71.54} 
            & \textcolor{gray}{69.59} 
            & \textcolor{gray}{77.40} 
            & \textcolor{gray}{-}
            \\
            
            \hdashline
            
            \textit{Direct}~\cite{hashizume_mix} 
            & 85.5      
            & 37.1      
            & 31.3      
            & 44.7      
            & 74.7
            \\
            
            ~~~w/ DA 
            & 97.93     
            & 68.22     
            & 69.22     
            & 63.24      
            & 89.99
            \\

            ~~~w/ DA, Reconst~(\textit{Direct-Reconst}) 
            & \bf 98.25 
            & \bf 77.74 
            & \bf 79.20 
            & \bf 82.47 
            & \bf 94.62
            \\
            
            \hline
        \end{tabular}
        }
\end{table}
\begin{table}[tb]
    \centering
    \caption{Evaluation results of the MSS accuracy for the output separated sound in \textit{Cascade}. SDR (Signal-to-Distortion Ratio) was used for the evaluation. The results of \textit{Cascade} w/ Spleeter~\cite{hashizume_inst} are quoted from the previous study~\cite{hashizume_inst}. Museval~\cite{museval} was used for calculation of SDR.}
    \label{table:ex_result_inst_sdr}
    \scalebox{0.85}[0.85]{
        \begin{tabular}{l|cccc}
            \hline
            & \multicolumn{4}{c}{SDR}\\
            Method 
            & drums 
            & bass 
            & piano 
            & guitar
            \\
            
            \hline
            
            \textit{Cascade} w/ Spleeter~\cite{hashizume_inst} 
            & -13.7      
            & -15.5      
            & -14.7     
            & - 
            \\
            
            \textit{Cascade} w/ from-scratch        
            & 15.50     
            & 10.54     
            & 7.81     
            & 6.94
            \\

            \hline
        \end{tabular}
    }
\end{table}

Evaluation results of MES-Normal and MES-Pseudo are shown in Table~\ref{table:ex_result_normal} and Table~\ref{table:ex_result_pseudo}, respectively. 
We also show an evaluation result of MSS accuracy for the output separated sounds in \textit{Cascade} methods in Table~\ref{table:ex_result_inst_sdr}.
Additionally, the results of perceptual similarity agreement are shown in Table \ref{table:ex_result_abx}. 

\subsubsection{Evaluation of \textit{Cascade-FT}}\label{sec: eval_cascade-ft}
It can be observed from Table~\ref{table:ex_result_normal} that \textit{Cascade-FT} achieves higher evaluation scores than the previous method~\cite{hashizume_inst} for all instruments. This suggests that the proposed methods achieve higher InMSRL performance compared to the previous method. From a comparison between \textit{Cascade} w/ from-scratch, w/o pseudo-musical-pieces and \textit{Cascade} w/ Spleeter~\cite{hashizume_inst}, the performance improvements can be seen in former.
This is caused by the insufficient separation accuracy of Spleeter, as shown in Table \ref{table:ex_result_inst_sdr}.
This poor performance of Spleeter is likely due to the fact that Spleeter is trained on music with raw-audio-songs, while the experiments in this paper and \cite{hashizume_inst} use musical pieces generated from MIDI.
Moreover, we can also observe that E2E-FT in the proposed method is effective for further performance improvements from a comparison between \textit{Cascade} w/ from-scratch and \textit{Cascade-FT}.
This results demonstrates that the performance of the MSS model in \textit{Cascade} methods strongly affects the accuracy of InMSRL.

The disentanglement performance of each InMSRL method can be compared in Table~\ref{table:ex_result_pseudo}.
All evaluation scores of \textit{Cascade-FT} exceed 90\%.
We also observe that E2E-FT is helpful to further improve the performance.

These results suggest that the proposed method \textit{Cascade-FT} can learn high-quality music similarity feature representations focusing on individual instrument sounds.

\subsubsection{Evaluation of \textit{Direct-Reconst}}\label{sec:eval_mix}
Table~\ref{table:ex_result_normal} shows that \textit{Direct-Reconst} does not outperform the previous method~\cite{hashizume_mix} for some instruments, i.e., bass, piano, and guitar.
On the other hand, \textit{Direct-Reconst} significantly outperforms previous method in the evaluation result of MES-Pseudo as shown in Table~\ref{table:ex_result_pseudo}. 
These results indicate that \textit{Direct}~\cite{hashizume_mix} for MES-Normal is significantly affected by the leakage of the other instrument sounds and its disentanglement performance is actually low. 
On the other hand, the proposed method \textit{Direct-Reconst} not only improves the evaluation scores of MES-Pseudo but also maintains the evaluation scores of MES-Normal at the same level as the previous method.
Therefore, the proposed method can achieve better InMSRL performance than the previous method.
Table \ref{table:ex_result_pseudo} also shows that DA significantly improves the MES-Pseudo score, demonstrating the effectiveness of DA.
Additionally, a comparison of \textit{Direct} w/ DA and \textit{Direct-Reconst} in Tables~\ref{table:ex_result_normal} and \ref{table:ex_result_pseudo} shows that the multi-task learning of the disentangled music similarity feature extraction and the reconstruction is effective for improving the InMSRL performance.

\subsubsection{The results of visualizing the music similarity feature vectors}
\begin{figure}[t]
    \centerline{
        \includegraphics[width=1.0\columnwidth]
        {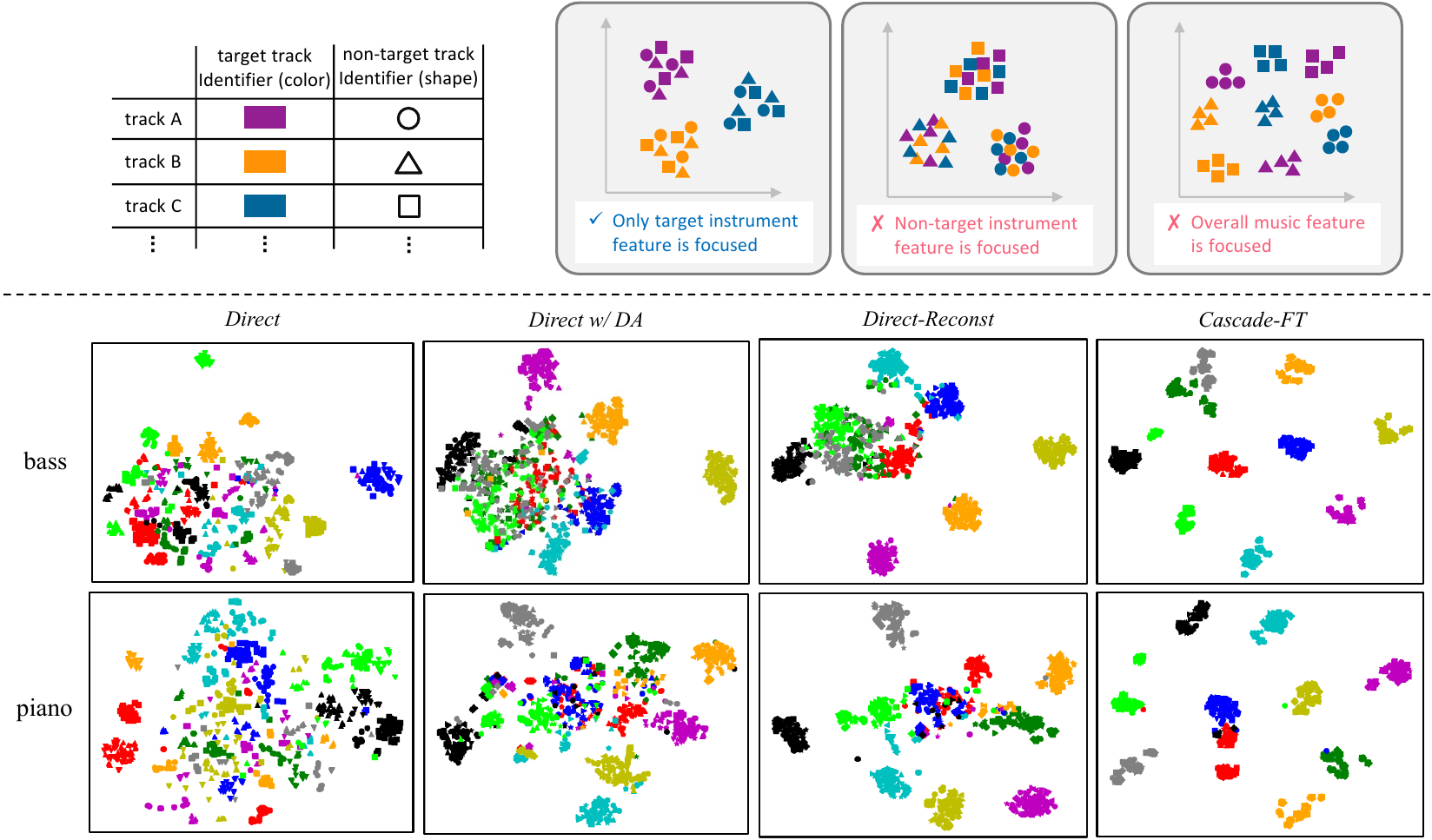}
        }
    \caption{Visualization results of the music similarity features for pseudo-musical-pieces. In visualization, the music identification for the target instrument is represented by colors, while that for non-target instruments is represented by shapes. In this setting, the aggregation of music similarity features with the same color but the different shapes indicates that the model focuses only on the feature of target instrument. In contrast, the aggregation of music similarity features with the same shape but different colors indicates that the model focuses on the features of non-target instrument, while the aggregation of music similarity features with the same shape and color indicates that the model focuses on the features of overall musical pieces. The music similarity feature vectors were compressed to 2 dimention vectors by t-SNE \cite{tsne}.}
    \label{fig:psd_mine}
\end{figure}

In Fig. \ref{fig:psd_mine}, the music similarity features values of \textit{Direct-Reconst} exhibit the most ideal aggregation pattern among the \textit{Direct}-retated approaches.
In the results of \textit{Direct} \cite{hashizume_mix}, clusters with the same color and shape are observed, indicating that the model focuses on the feature of overall musical pieces.
In addition, a dispersion of feature values is seen in the results of \textit{Direct} \cite{hashizume_mix}, suggesting that the performance of InMSRL and disentanglement is insufficient.
Additionally, in the \textit{Direct w/ DA}, clusters of the same shape but different shapes are observed, indicating that the model focuses on the features of non-target instruments and is also inadequate at disentangling individual instrument features.
The clusters in \textit{Direct-Reconst} are mainly composed of the same color but different shapes, and they also exhibit reduced feature dispersion compared to \textit{Direct} and \textit{Direct w/ DA}.
This indicates that \textit{Direct-Reconst} achieves higher disentanglement performance and improved InMSRL performance compared to conventional methods, demonstrating the effectiveness of the multi-task learning approach for \textit{Direct}.
Furthermore, the results of \textit{Cascade-FT} show sufficiently reduced feature dispersion and successfully form clusters with the same color but different shapes.

\subsubsection{Evaluation of perceptual InMSRL performance}
\begin{table}[tb]
    \centering
    \caption{Evaluation results of perceptual similarity agreement. The ``w/o PAFT'' columns show the scores of models only trained with S4 similarity, and The ``w/ PAFT'' columns show the scores of models trained with PAFT. $[*, *]$ represents the 95\% confidence interval calculated using the Clopper-Pearson method \cite{cp}, and $\underset{\pm *}{*}$ represents the mean and standard deviation from three training runs, highlighting the instability in the model’s behavior caused by the limited data available for PAFT. The ``mean'' columns indicates the average evaluation score of drums, bass, piano, and guitar. The ``PAFT data'' column indicates the input data of models during PAFT. Bold indicates the highest value, while the underline represents the second-highest value. ``\textcolor{gray}{Gray}'' text indicates the scores used in the ablation study.}
    \label{table:ex_result_abx}
    \scalebox{0.615}[0.615]{
        \begin{tabular}{lc||cccccc|cccccc}
            \hline
            \multirow{2}{*}{\makecell[l]{Method}}
            & \multirow{2}{*}{\makecell[c]{PAFT\\data}}
            & \multicolumn{6}{c}{\bf w/o PAFT}
            & \multicolumn{6}{c}{\bf w/ PAFT}
            \\ 

            &
            & drums
            & bass
            & piano
            & guitar
            & residuals
            & mean
            & drums
            & bass
            & piano
            & guitar
            & residuals
            & mean
            \\
            
            \hline

            \rowcolor{lightgray!50} All-Diff &&&&&&&&&&&&&\\
            

            \textit{Clean} 
            & clean
            & $\underset{[60.27, 65.89]}{63.11}$
            & $\underset{[52.18, 58.07]}{55.14}$
            & $\underset{[58.62, 64.50]}{61.59}$
            & $\underset{[62.62, 68.36]}{\underline{65.53}}$
            & $-$
            & $61.34$
            & $\underset{\pm 5.62}{61.81}$
            & $\underset{\pm 17.20}{71.89}$
            & $\underset{\pm 2.64}{58.36}$
            & $\underset{\pm 3.67}{66.10}$
            & $-$
            & $64.54$
            \\

            \hdashline

            \multirow{2}{*}{\textit{Cascade}}
            & pseudo
            & \multirow{3}{*}{$\underset{[66.83, 72.20]}{\bf 69.56}$}
            & \multirow{3}{*}{$\underset{[59.42, 65.16]}{\underline{62.32}}$}
            & \multirow{3}{*}{$\underset{[54.33, 60.31]}{57.34}$}
            & \multirow{3}{*}{$\underset{[63.84, 69.53]}{\bf 66.73}$}
            & \multirow{3}{*}{$-$}
            & \multirow{3}{*}{$\bf 63.99$}
            & $\underset{\pm 0.75}{\bf 71.09}$
            & $\underset{\pm 2.58}{\underline{75.50}}$
            & $\underset{\pm 5.52}{60.45}$
            & $\underset{\pm 4.62}{\bf 76.55}$
            & $-$
            & $\underline{70.90}$
            \\

            & \textcolor{gray}{normal}
            &&&&&&
            & \textcolor{gray}{$\underset{\pm 3.07}{69.53}$}
            & \textcolor{gray}{$\underset{\pm 5.42}{73.81}$}
            & \textcolor{gray}{$\underset{\pm 0.80}{\underline{64.22}}$}
            & \textcolor{gray}{$\underset{\pm 2.72}{62.99}$}
            & \textcolor{gray}{$-$}
            & \textcolor{gray}{$67.64$}
            \\

            \textit{Cascade-PAFT} 
            & pseudo
            &&&&&&
            & $\underset{\pm 3.49}{\underline{70.54}}$
            & $\underset{\pm 4.60}{\bf 76.30}$
            & $\underset{\pm 4.62}{\bf 64.65}$
            & $\underset{\pm 0.16}{\underline{75.80}}$
            & $-$
            & $\bf 71.83$
            \\

            \hdashline
            
            \multirow{2}{*}{\textit{Cascade-FT}} 
            & pseudo
            & \multirow{2}{*}{$\underset{[59.40, 65.05]}{\underline{62.25}}$}
            & \multirow{2}{*}{$\underset{[58.52, 64.29]}{61.44}$}
            & \multirow{2}{*}{$\underset{[59.46, 65.31]}{\underline{62.42}}$}
            & \multirow{2}{*}{$\underset{[55.53, 61.48]}{58.53}$}
            & \multirow{2}{*}{$-$}
            & \multirow{2}{*}{$61.16$}
            & $\underset{\pm 2.91}{69.69}$
            & $\underset{\pm 3.70}{68.75}$
            & $\underset{\pm 3.68}{59.75}$
            & $\underset{\pm 4.90}{73.45}$
            & $-$
            & $67.91$
            \\

            & \textcolor{gray}{normal}
            &&&&&&
            & \textcolor{gray}{$\underset{\pm 4.26}{69.77}$}
            & \textcolor{gray}{$\underset{\pm 2.77}{71.65}$}
            & \textcolor{gray}{$\underset{\pm 2.64}{60.19}$}
            & \textcolor{gray}{$\underset{\pm 0.59}{64.59}$}
            & \textcolor{gray}{$-$}
            & \textcolor{gray}{$66.55$}
            \\

            \hdashline

            \textit{Direct} 
            & pseudo
            & $\underset{[53.41, 59.19]}{56.74}$
            & $\underset{[62.93, 68.55]}{\bf 65.78}$
            & $\underset{[57.12, 63.04]}{\bf 69.64}$
            & $\underset{[59.72, 65.56]}{62.20}$
            & $\underset{[62.63, 68.29]}{57.63}$
            & $\underline{61.55}$
            & $\underset{\pm 0.80}{63.95}$
            & $\underset{\pm 2.24}{68.51}$
            & $\underset{\pm 1.67}{62.64}$
            & $\underset{\pm 2.37}{64.50}$
            & $\underset{\pm 3.14}{\underline{57.11}}$
            & $64.90$
            \\
            
            \textit{Direct-Reconst} 
            & pseudo
            & $\underset{[53.41, 59.19]}{56.32}$
            & $\underset{[62.93, 68.55]}{\bf 65.78}$
            & $\underset{[57.12, 63.04]}{60.11}$
            & $\underset{[59.72, 65.56]}{62.67}$
            & $\underset{[62.63, 68.29]}{\bf 65.50}$
            & $61.47$
            & $\underset{\pm 1.64}{55.27}$
            & $\underset{\pm 3.69}{72.93}$
            & $\underset{\pm 1.06}{56.08}$
            & $\underset{\pm 2.55}{71.94}$
            & $\underset{\pm 3.39}{\bf 62.79}$
            & $64.06$
            \\
            
            \hline

            \rowcolor{lightgray!50} One-Shared &&&&&&&&&&&&&\\
            

            \textit{Clean} 
            & clean
            & $\underset{[94.35, 96.18]}{95.33}$
            & $\underset{[91.13, 93.53]}{\bf 92.39}$
            & $\underset{[93.24, 95.24]}{\underline{94.30}}$
            & $\underset{[93.06, 95.10]}{\bf 94.14}$
            & $-$
            & $\bf 94.04$
            & $\underset{\pm 0.00}{\bf 95.90}$
            & $\underset{\pm 1.20}{\bf 92.43}$
            & $\underset{\pm 0.64}{\underline{92.66}}$
            & $\underset{\pm 0.77}{\underline{93.84}}$
            & $-$
            & $\bf 93.71$
            \\

            \hdashline

            \multirow{2}{*}{\textit{Cascade}} 
            & pseudo
            & \multirow{3}{*}{$\underset{[95.37, 97.02]}{\bf 96.26}$}
            & \multirow{3}{*}{$\underset{[89.82, 92.39]}{91.17}$}
            & \multirow{3}{*}{$\underset{[93.44, 95.41]}{\bf 94.49}$}
            & \multirow{3}{*}{$\underset{[91.95, 94.15]}{93.11}$}
            & \multirow{3}{*}{$-$}
            & \multirow{3}{*}{$93.76$}
            & $\underset{\pm 1.11}{95.26}$
            & $\underset{\pm 3.05}{89.71}$
            & $\underset{\pm 2.46}{92.49}$
            & $\underset{\pm 1.38}{92.13}$
            & $-$
            & $92.39$
            \\

            & \textcolor{gray}{normal}
            &&&&&&
            & \textcolor{gray}{$\underset{\pm 0.55}{\underline{95.58}}$}
            & \textcolor{gray}{$\underset{\pm 1.67}{\underline{92.30}}$}
            & \textcolor{gray}{$\underset{\pm 0.00}{94.56}$}
            & \textcolor{gray}{$\underset{\pm 1.22}{92.80}$}
            & \textcolor{gray}{$-$}
            & \textcolor{gray}{$93.81$}
            \\

            \textit{Cascade-PAFT} 
            & pseudo
            &&&&&&
            & $\underset{\pm 0.00}{\bf 95.90}$
            & $\underset{\pm 0.40}{90.13}$
            & $\underset{\pm 1.57}{93.13}$
            & $\underset{\pm 1.32}{91.18}$
            & $-$
            & $92.59$
            \\

            \hdashline
            
            \multirow{2}{*}{\textit{Cascade-FT}} 
            & pseudo
            & \multirow{2}{*}{$\underset{[95.37, 97.02]}{\bf 96.26}$}
            & \multirow{2}{*}{$\underset{[90.75, 93.20]}{\underline{92.04}}$}
            & \multirow{2}{*}{$\underset{[93.04, 95.07]}{94.12}$}
            & \multirow{2}{*}{$\underset{[92.15, 94.32]}{\underline{93.30}}$}
            & \multirow{2}{*}{$-$}
            & \multirow{2}{*}{$\underline{93.93}$}
            & $\underset{\pm 0.00}{\bf 95.90}$
            & $\underset{\pm 1.17}{90.93}$
            & $\underset{\pm 2.11}{\bf 92.70}$
            & $\underset{\pm 0.68}{\bf 93.97}$
            & $-$
            & $\underline{93.37}$
            \\

            & \textcolor{gray}{normal}
            &&&&&&
            & \textcolor{gray}{$\underset{\pm 0.55}{\underline{95.58}}$}
            & \textcolor{gray}{$\underset{\pm 0.40}{92.01}$}
            & \textcolor{gray}{$\underset{\pm 1.57}{93.35}$}
            & \textcolor{gray}{$\underset{\pm 1.32}{90.33}$}
            & \textcolor{gray}{$-$}
            & \textcolor{gray}{$92.82$}
            \\

            \hdashline

            \textit{Direct} 
            & pseudo
            & $\underset{[94.19, 97.21]}{\underline{95.90}}$
            & $\underset{[89.09, 93.36]}{91.40}$
            & $\underset{[84.89, 89.69]}{87.44}$
            & $\underset{[86.74, 91.35]}{89.20}$
            & $\underset{[90.83, 94.86]}{\bf 93.04}$
            & $90.99$
            & $\underset{\pm 0.00}{\bf 95.90}$
            & $\underset{\pm 0.48}{92.10}$
            & $\underset{\pm 1.05}{86.10}$
            & $\underset{\pm 0.94}{90.96}$
            & $\underset{\pm 0.00}{90.62}$
            & $91.26$
            \\
            
            \textit{Direct-Reconst}
            & pseudo
            & $\underset{[95.01, 96.72]}{95.93}$
            & $\underset{[87.72, 90.52]}{89.18}$
            & $\underset{[90.17, 92.58]}{91.43}$
            & $\underset{[91.35, 93.63]}{92.55}$
            & $\underset{[90.80, 93.24]}{92.09}$
            & $92.24$
            & $\underset{\pm 1.31}{93.57}$
            & $\underset{\pm 2.82}{84.91}$
            & $\underset{\pm 2.59}{86.40}$
            & $\underset{\pm 0.90}{88.98}$
            & $\underset{\pm 1.03}{86.18}$
            & $88.46$
            \\

            \hline
        \end{tabular}
        }
\end{table}
Table \ref{table:ex_result_abx} shows the evaluation results of perceptual InMSRL performance.
\textit{Cascade} in Table \ref{table:ex_result_abx} is corresponding to \textit{Cascade} w/ from-scratch in Table \ref{table:ex_result_normal} and \ref{table:ex_result_pseudo}, and \textit{Direct} in Table \ref{table:ex_result_abx} is corresponding to \textit{Direct} w/ DA in Table \ref{table:ex_result_normal} and \ref{table:ex_result_pseudo}.
Moreover, \textit{Cascade-PAFT} refers to the models that replaces the E2E-FT of \textit{Cascade-FT} with PAFT and conducts the E2E-FT and PAFT simultaneously.

First, we discuss the perceptual InMSRL performance of models without PAFT as shown in ``w/o PAFT'' columns of Table \ref{table:ex_result_abx}.
Under the One-Shared condition, \textit{Cascade-FT} improves perceptual InMSRL performance compared to \textit{Cascade}, and \textit{Direct-Reconst} improves perceptual InMSRL performance compared to \textit{Direct}.
This is expected since the One-Shared condition can be regarded as based on S4, indicating that the proposed methods achieve better perceptual InMSRL performance between the same musical pieces than previous methods.

Furthermore, in ``w/o PAFT'' columns of Table \ref{table:ex_result_abx}, the scores of models under the All-Diff condition are significantly lower than the scores of models under the One-Shared condition.
However, this does not mean that similarity is not represented at all in the All-Diff condition, i.e., all models achieved an average perceptual similarity agreement performance of over 60\%.
This results are similar to the previous study \cite{abx}.
These results indicate that S4-based training contributes to some extent not only to capturing similarity within same musical pieces but also to representing similarity between different musical pieces although this contribution is not sufficient.

Moreover, under the All-Diff condition, \textit{Cascade} outperforms \textit{Cascade-FT} and \textit{Direct} outperforms \textit{Direct-Reconst}, which is the opposite result of the One-Shared condition.
This means that effective learning under the training strategy based on S4 leads to higher InMSRL performance between the same musical pieces but does not contribute to improving InMSRL performance between different musical pieces.

Next, we discuss the perceptual InMSRL performance of models with PAFT as shown in ``w/ PAFT'' columns of Table \ref{table:ex_result_abx}.
We can observe that scores of models with PAFT are higher than those of models without PAFT under the All-Diff condition. 
In contrast, the scores under the One-Shared condition remain at the same performance level.
These results demostrate that PAFT contributes not only to enhancing the perceptual InMSRL performance between different musical pieces, which cannot be sufficiently trained based on S4 similarity, but also to maintaining to represent the similarity between same musical pieces.

Additionally, when comparing to the perceptual InMSRL performance between different musical pieces of models with PAFT, \textit{Cascade} (row 2) is superior to \textit{Cascade-FT} (row 5), and \textit{Direct} (row 7) is superior to \textit{Direct-Reconst} (row 8).
This indicates that an effective training strategy based on S4 similarity does not necessarily lead to performance improvement through PAFT.
It can also be considered that there is a significant discrepancy between the features emphasized across different musical pieces and those emphasized within the same musical piece.

Furthermore, \textit{Cascade-PAFT} achieves the highest perceptual InMSRL performance under All-Diff condition.
This can be considered as a result not only of minimizing the impact of separation errors from MSS model on the subsequent feature extractor but also of optimizing the entire network for music similarity based on human preference.
This indicates that E2E-FT is also effective for improving perceptual InMSRL performance.

\subsubsection{\textit{Clean}, \textit{Cascade} and \textit{Direct} Approaches Comparison}
First, we compare \textit{Cascade} approach with \textit{Direct} approach.
Tables \ref{table:ex_result_normal} and \ref{table:ex_result_pseudo} show that \textit{Direct} approach tends to have higher MES-Normal scores than MES-Pseudo scores for some instruments except for drums and residuals.
Normally, MES-Normal score would be lower than or equal to the MES-Pseudo score because the MES-Normal uses 136 target labels compared to 10 for the MES-Pseudo at 5NN.
This is considered to be due to the leakage of the other instrument sounds as discussed in Section \ref{sec:eval_mix}.
In contrast, the \textit{Cascade} approach can more precisely focus only on target instrument sound, as demonstrated by the higher MES-Pseudo scores than the MES-Normal scores.
Additionally, in the visualization results from Fig. \ref{fig:psd_mine}, \textit{Cascade-FT} surpresses feature dispersion and successfully forms clusters with the same color but different shapes compared to \textit{Direct} approach.
These results demonstrate that \textit{Cascade} approach achieves higher InMSRL performance and disentanglement performance than \textit{Direct} approach.
Furthermore, from All-Diff scores of ``w/ PAFT'' columns of Table \ref{table:ex_result_abx}, \textit{Cascade} approach also tends to have higher perceptual InMSRL performance between different musical pieces than \textit{Direct} approach.
This indicates that \textit{Cascade} approach also outperforms \textit{Direct} approach in terms of the effectiveness of PAFT.
These several results demonstrates that \textit{Cascade} approach achieves higher InMSRL performance not only in objective evaluation but also in perceptual similarity representation than \textit{Direct} approach.
On the other hand, \textit{Direct} approach needs to use only the disentangled music similarity feature extractor in the inference step, and therefore, its computational cost is lower than \textit{Cascade} approach.

Next, we compare \textit{Clean} approach with \textit{Cascade} and \textit{Direct} approach.
MES-Normal scores of \textit{Clean} are the most highest in all of InMSRL models as shown in Table \ref{table:ex_result_normal} and it is predicted that MES-Pseudo scores would be much higher scores than MES-Normal scores because of its less target labels.
Additionally, \textit{Clean} records the highest scores on the perceptual similarity agreement under the One-Shared condition as shown in ``w/o PAFT'' columns of  Table \ref{table:ex_result_abx}.
This result is to be expected because \textit{Clean} utilizes clean individual instrument sounds as input, which are generally not publicly available, therefore explicitly providing distinct individual instrument features to the music similarity feature extractors.
In contrast, in terms of perceptual InMSRL performance between different musical pieces, \textit{Clean} is not the top-performing model.
Specifically, the scores of \textit{Clean} without PAFT under the All-Diff condition are comparable to those of the \textit{Cascade} and \textit{Direct} approaches without PAFT as shown in ``w/o PAFT'' columns of Table \ref{table:ex_result_abx}, and the \textit{Cascade} approach with PAFT rather outperforms \textit{Clean} with PAFT as shown in ``w/ PAFT'' columns of Table \ref{table:ex_result_abx}.
It is possible that this is caused by the overlearning of the music similarity between the same musical pieces.
Since learning the similarity between the same musical pieces possibly requires only a specific part of the instrument features, such as hi-hat features in drums, models that use clean instrument sounds have potential to focus primarily on those specific features.
In such cases, the model is expected to be struggle to perform well when it needs to capture more diverse features, such as when learning similarity between different musical pieces.
Conversely, in \textit{Cascade}, while the feature extractors receive separated instrument sounds that contain separation errors, these errors possibly mask some parts of the instrument features and potentially lead to more robust similarity representation learning.
As a result, in perceptual InMSRL between different musical pieces with PAFT, the model was able to capture various features, which likely contributed to performance improvement.

From evaluation results, \textit{Clean} is superior to \textit{Cascade} and \textit{Direct} for the InMSRL performance between same musical pieces, while \textit{Cascade} tends to outperform \textit{Clean} for the InMSRL performance between different musical pieces.

\subsubsection{The effectiveness of pseudo-musical-pieces}\label{sec: ex_result_pseudo}
In \textit{Cascade} approach, the evaluation result of w/o pseudo-musical-pieces showed in Table~\ref{table:ex_result_normal} and Table~\ref{table:ex_result_pseudo} indicates that by using pseudo-musical-pieces, we can minimize the adverse effects of separation errors caused by the MSS model. 
Note that although the performance w/o pseudo-musical-pieces looks higher than that w/ it in Table \ref{table:ex_result_normal}, this result is caused by the leakage of the other instrument sounds, and therefore, the actual InMSRL performance is limited. 
The use of pseudo-musical-pieces is also essential in \textit{Direct} approach as reported in \cite{hashizume_mix}. 
These results demonstrate that the use of pseudo-musical-pieces is an important technique to improve InMSRL performance.
Furthermore, in Table \ref{table:ex_result_abx}, the All-Diff scores of \textit{Cascade} approach which utilizes pseudo-musical-pieces during PAFT are higher than those of \textit{Cascade} approach which utilizes normal musical pieces during PAFT.
This means that pseudo-musical-pieces are more effective than normal musical pieces during PAFT in terms of perceptual InMSRL performance between different musical pieces since they minimizes distraction from the features of other instrument sounds.

\section{Conclusion}\label{sec: conclusion}
In this paper, we have proposed three methods to improve InMSRL performance. 
First, for \textit{Cascade}, we have proposed end-to-end fine-tuning (E2E-FT) of the MSS model and the music similarity feature extractors using an auxiliary separation loss.
Second, for \textit{Direct}, we have proposed joint training of the disentangled feature extraction and MSS based on the reconstruction with the disentangled music similarity features.
Third, we employ perception-aware fine-tuning (PAFT) utilizing human preference.
We have conducted experimental evaluations and have demonstrated that the E2E-FT for \textit{Cascade} improves InMSRL performance, the multi-task learning for \textit{Direct} is also helpful to improve disentanglement performance in the feature extraction, PAFT enhances the perceptual InMSRL performance, and \textit{Cascade} with the E2E-FT and PAFT outperforms \textit{Direct} with the multi-task learning and PAFT.
Future work includes using raw-audio-songs and vocal sound.



\printbibliography

\end{document}